\documentclass[rnote]{aa}  
\usepackage[pdftex]{color,xcolor}

\begin{document} 
  \title{Astrometric mass ratios for three spectroscopic binaries } 
\author{J. Sahlmann\inst{1}
		\and F. C. Fekel\inst{2}}		
\institute{Observatoire de Gen\`eve, Universit\'e de Gen\`eve, 51 Chemin Des Maillettes, 1290 Versoix, Switzerland\\
		\email{johannes.sahlmann@unige.ch}
		\and	
		Center of Excellence in Information Systems, Tennessee State University, 3500 John A. Merritt Boulevard, Box 9501, Nashville, TN 37209, USA 	\\
\email{fekel@evans.tsuniv.edu}	      }
			
\date{Received 13 June 2013 / Accepted 10 July 2013} 

\abstract
{The orbits of five single-lined spectroscopic binaries have recently been determined. We now
use astrometric measurements that 
were collected with the Hipparcos satellite to constrain the systems' mass 
ratios and secondary masses. The barycentric astrometric orbits of three 
binary systems, HD\,140667, HD\,158222, and HD\,217924, are fully 
determined and precise estimates of their mass ratios are obtained. 
Follow-up of these systems with infrared spectroscopy could yield 
model-independent dynamical masses for all components.}

\maketitle
\section{Introduction}
Spectroscopic binaries make it possible to study stellar multiplicity over 
a wide range of secondary masses and are therefore one of the foundations 
for our understanding of star and binary formation \citep{Duquennoy:1991kx, 
Raghavan:2010fk}. The orbital elements of a single-lined spectroscopic 
binary (SB1) produce the mass function $f(m)$, whose value depends on the 
primary mass $M_1$, the secondary mass $M_2$, and the inclination of the 
orbital plane $i$ relative to the plane of the sky, which usually is 
unknown. If an estimate of $M_1$ can be made, the minimum mass of the
secondary, $M_2 \sin i$, is obtained. Therefore, only a lower limit to 
the secondary-to-primary mass ratio $q=M_2/M_1$ is known for most SB1s. 
Astrometric measurements of orbital motion can determine the inclination, 
and thus the mass ratio. Obtaining a large number of systems with 
well-determined $q$ values helps to refine our knowledge of the binary 
population (\citealt{Tohline:2002lr, Goodwin:2013fk}) and SB1s with 
astrometric orbits (e.g. \citealt{Pourbaix:2000sf, Jancart:2005mz,  Ren:2013fk}) can 
complement the samples of double-lined spectroscopic binaries (SB2, e.g. 
\citealt{Mazeh:2003fe}) and  eclipsing binaries. Here, we study 
five new SB1s analyzed by \cite{Fekel:2013fk} that have primary spectral 
types ranging from F9~V to G5~V and whose basic properties are given in 
Table\,\ref{tab:1}. These binaries have eccentricities $e\approx0.2-0.84$, 
orbital periods $P\approx60-2400$ days, and radial velocity (RV) 
amplitudes of $K_1\approx5-27$ km/s.

\section{Primary mass estimates}
\begin{figure}
\begin{center} 
\includegraphics[width= 0.70\linewidth]{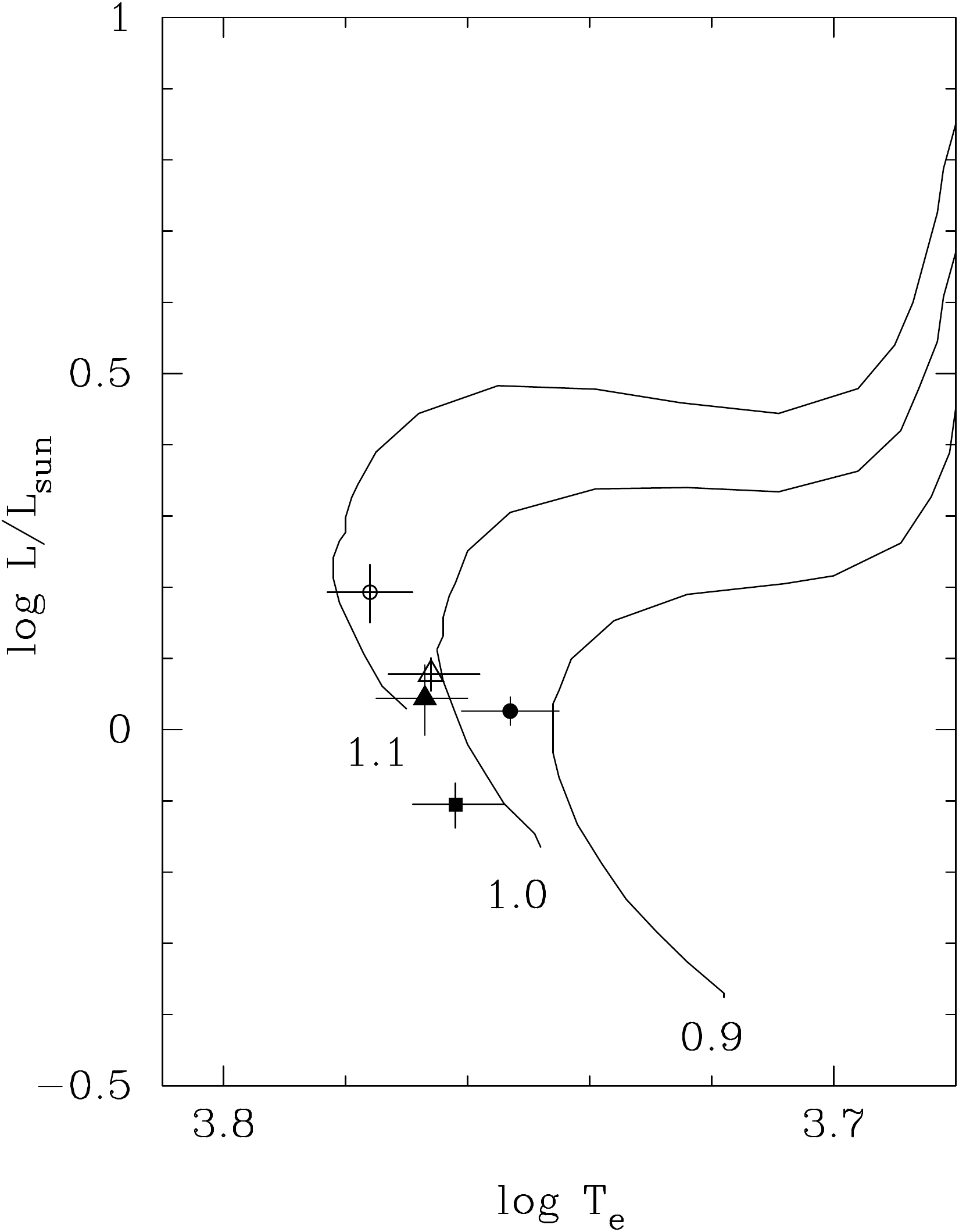} 
\caption{Hertzsprung-Russell diagram showing the locations of HD\ 100167 
(open triangle), HD\ 135991 (open circle), HD\ 140667 (filled triangle), 
HD\ 158222 (filled square), and HD\ 217924 (filled circle). Theoretical 
evolutionary tracks of solar composition are labelled for masses of 0.9, 
1.0, and 1.1 $M_{\sun}$. \vspace{-0.6cm}}
\label{fig:HR}
\end{center} \end{figure}

Estimates of the primary masses were obtained by determining the effective 
temperature and luminosity of each star and then comparing those results 
with theoretical evolutionary tracks. For each system we began by adopting 
the $V$ mag and $B-V$ colour from the Hipparcos catalog \citep{Perryman:1997kx}. 
We next adopted the parallax from our revised astrometric solution, if 
available (see Sect.~\ref{sec:hip}), or the parallax reduction by 
\cite{:2007kx}. The apparent magnitude and parallax result in the absolute 
magnitude. The $B-V$ colour in conjunction with Table 3 of \cite{Flower:1996qy} provides the bolometric correction and effective temperature for each star. 
Converting the bolometric magnitude to luminosity in solar units, we then 
plotted the stars in a theoretical H-R diagram (Fig.~\ref{fig:HR}) and 
compared them with the solar abundance evolutionary tracks of 
\cite{Girardi:2000bf}. The primary mass estimates are listed in 
Table~\ref{tab:1} with the corresponding minimum secondary masses.

\begin{table}
\caption{Primary masses and minimum secondary masses of the five systems. 
Mass function, period, and eccentricity are given for orientation, see exact 
values in \cite{Fekel:2013fk}.}
\label{tab:1} 
\centering  
\begin{tabular}{c r r r r r r r} 
\hline\hline %
HD        &$f(m)$           & $M_1$           & $M_2 \sin i$ &  $P$ & $e$\\
            &($M_{\sun}$) & ($M_{\sun}$) &($M_{\sun}$)& (day)& \\
\hline 
100167 & 0.0489 & $1.01$ & $0.48$  & $ 60.6$ & $0.68$ \\
135991 & 0.0045 & $1.07$ & $0.19$  & $151.0$ & $0.57$ \\
140667 & 0.0100 & $1.04$ & $0.26$  & $978.4$ & $0.20$ \\
158222 & 0.0214 & $0.94$ & $0.32$  & $206.1$ & $0.41$ \\
217924 & 0.0119 & $1.05$ & $0.28$  & $2402.7$ & $0.84$ \\
\hline
\end{tabular} 
\end{table}

\section{Analysis of the Hipparcos astrometry}\label{sec:hip}
All stars listed in Table~\ref{tab:1} were catalogued by the Hipparcos 
astrometry satellite \citep{Perryman:1997kx}. We used the new Hipparcos 
reduction \citep{:2007kx} to search for signatures of orbital motion in 
the Intermediate Astrometric Data (IAD). The analysis was performed as 
described in \cite{Sahlmann:2011fk}, where a detailed description of 
the method can be found. The main reduction elements are as follows. 
Using the spectroscopic orbital parameters of \cite{Fekel:2013fk}, the 
IAD was fitted with a seven-parameter model, where the free parameters 
are the inclination $i$, the longitude of the ascending node $\Omega$, 
the parallax $\varpi$, and offsets to the coordinates ($\Delta \alpha^{\star}$, 
$\Delta \delta$) and proper motions ($\Delta \mu_{\alpha^\star}$, 
$\Delta \mu_{\delta}$). A two-dimensional grid in $i$ and $\Omega$ 
(see Fig.~\ref{fig:contour}) was searched for its global $\chi^2$-minimum 
with a standard nonlinear minimisation procedure. The statistical 
significance of the derived astrometric orbit was determined with a 
permutation test employing 1000 pseudo orbits. Uncertainties in the 
solution parameters were derived by Monte Carlo simulations that include 
propagation of RV parameter uncertainties. This method has proven to be 
reliable in detecting orbital signatures in the Hipparcos IAD 
\citep{Sahlmann:2011fk, Sahlmann:2011lr,Diaz:2012fk}. Because the binaries 
are SB1s, it is reasonable to assume that the secondaries' light has a 
negligible effect on the Hipparcos astrometry, i.e.\ the photocentric and 
barycentric positions coincide. Table~\ref{tab:2} gives the new Hipparcos 
catalogue parameters of the five primaries. HD\,217924 and HD\,140667 have 
a stochastic ('1') and accelerated ('7') solution type, respectively, which 
is common for systems with unrecognised orbital motion \citep{Sahlmann:2011fk}. 
The remaining three systems have standard five-parameter solutions ('5'). 
The parameter $N_\mathrm{orb}$ represents the number of orbital periods 
covered by the Hipparcos observation timespan and $N_\mathrm{Hip}$ is the 
number of IAD measurements in \cite{:2007kx} subtracted by the number of 
measurements $N_\mathrm{rej}$ that we rejected. The median astrometric 
precision is given by $\sigma_{\Lambda}$.

\begin{table}
\caption{Characteristics of the new reduction Hipparcos IAD}
\small
\label{tab:2} 
\centering  
\begin{tabular}{c c r r r r r r r r r r} 
\hline\hline %
HD & HIP & Sol. & $N_\mathrm{orb}$ & $\sigma_{\Lambda}$  & $N_\mathrm{Hip}$ & $N_\mathrm{rej}$\\  
      &     & type &           & (mas)                 &    &       \\  
\hline 
100167 & 056257 & 5 & 17.5 & 2.7 & 96  &0\\ 
135991 & 074821 & 5 & 6.5 & 4.3 & 124  &1\\ 
140667 & 077098 & 7 & 0.8 & 2.8 & 48  &4\\ 
158222 & 085244 & 5 & 5.6 & 3.5 & 102  &1\\ 
217924 & 113884 & 1 & 0.5 & 5.2 & 48  &0\\ 
 \hline
\end{tabular} 
\end{table}

\section{Results}
We performed the analysis for the five systems and found that the orbits 
of HD\,140667, HD\,158222, and HD\,217924 are constrained by the Hipparcos 
astrometry. Figures~\ref{fig:contour} and \ref{fig:orbits} show the 
corresponding confidence contours and the barycentric orbits, respectively. 
Tables~\ref{tab:3} and \ref{tab:4} summarise the numerical results. The 
orbital motion of HD\,100167 and HD\,135991 is not detected. The systems 
are discussed individually below. 
\begin{table*}
\caption{Astrometric solution parameters for the three significant orbits}
\small
\label{tab:3} 
\centering  
\begin{tabular}{c r r r r r r r r r} 
\hline\hline %
HD  & $\Delta \alpha^{\star}$ & $\Delta \delta$ & $\varpi$ &$\Delta \varpi$  & $\Delta \mu_{\alpha^\star}$ & $\Delta \mu_{\delta}$ & $i$ & $\Omega$  \\  
       &  (mas)                     & (mas)             &  (mas)    &  (mas)            &  (mas $\mathrm{yr}^{-1}$)       & (mas $\mathrm{yr}^{-1}$) & (deg)	      & (deg)	    \\  
\hline 
140667 & $-2.7^{+ 0.9}_{-0.9}$ & $-2.9^{+ 0.7}_{-0.7}$  & $27.0^{+ 1.4}_{-1.5}$ & $-3.4$ & $-7.5^{+ 2.2}_{-2.2}$ & $-3.5^{+ 3.2}_{-3.5}$ & $ 113.3^{+ 6.0}_{-6.5}$ & $ 69.9^{+ 13.8}_{-14.7}$  \vspace{1mm} \\ 
158222 & $-1.6^{+ 0.5}_{-0.5}$ & $0.3^{+ 0.5}_{-0.5}$  & $24.5^{+ 0.5}_{-0.5}$ & $0.5$ & $-1.9^{+ 0.5}_{-0.5}$ & $-1.5^{+ 0.6}_{-0.6}$ & $ 62.4^{+ 5.6}_{-5.1}$ & $ 221.1^{+ 6.8}_{-6.8}$  \vspace{1mm} \\ 
217924 & $-32.2^{+ 3.8}_{-3.7}$ & $-4.3^{+ 9.8}_{-9.2}$  & $37.2^{+ 1.3}_{-1.2}$ & $-1.2$ & $-17.0^{+ 1.7}_{-1.7}$ & $0.2^{+ 3.8}_{-3.8}$ & $ 51.3^{+ 4.6}_{-4.0}$ & $ 233.4^{+ 18.2}_{-16.5}$  \vspace{0.5mm} \\ 
 \hline
\end{tabular} 
\end{table*}

\begin{table*}
\caption{Astrometric solution parameters. Some parameters are omitted for the two systems with non-detected orbits.}
\small
\label{tab:4} 
\centering  
\begin{tabular}{c r r r r r r r r r r r} 
\hline\hline %
HD &   $a \sin i$ & $a$ & $M_2$  & $M_2$ (3-$\sigma$)& $a_{\mathrm{rel}}$ &O-C$_5$ &O-C$_7$ & $\chi^2_{7,red}$ & Null prob. & Significance \\  
      &   (mas)      &(mas)   & ($M_\sun$) & ($M_\sun$) &(mas) & (mas) & (mas) & & (\%)  & (\%) \\  
\hline 
100167 &   3.12 & $\cdots$& $\cdots$ & $\cdots$& $\cdots$ & 3.52 & 3.40& 1.35& 1.4 &12.9 \vspace{1mm} \\ 
135991 &   1.80 & $\cdots$ & $\cdots$& $\cdots$ & $\cdots$ & 5.53 & 5.47& 1.32& 9.4 &6.2 \vspace{1mm} \\ 
140667 &   12.64 & $ 12.3^{+ 0.4}_{-0.4}$ & $ 0.29^{+ 0.02}_{-0.02}$ & $(0.26,0.36)$ & 57 & 4.57 & 2.88& 1.06& 5.6e-08 &12.0 \vspace{1mm} \\ 
158222 &   4.56 & $ 5.3^{+ 0.3}_{-0.3}$ & $ 0.38^{+ 0.02}_{-0.02}$ & $(0.33,0.46)$ & 18 & 4.58 & 4.04& 0.93& 1.0e-07 &99.3 \vspace{1mm} \\ 
217924 &   30.72 & $ 38.1^{+ 1.8}_{-1.8}$ & $ 0.37^{+ 0.03}_{-0.03}$ & $(0.31,0.48)$ & 146 & 6.61 & 2.71& 0.23& 1.6e-17 &$>99.9$ \vspace{0.5mm} \\ 
\hline
\end{tabular} 
\end{table*}

\subsection{HD\,217924}
The astrometric orbit is clearly detected with a significance of $>$$99.9$\% 
($>3.3\sigma$), which is also reflected in the small residuals of the 
7-parameter model (O-C$_7$) including the orbit compared to the standard 
5-parameter model residuals (O-C$_5$) and the corresponding small F-test 
null probability of the simpler model being true. The orbital signature is 
large with a barycentric semimajor axis of $a_1=38.1 \pm 1.8$~mas. As a
consequence, the orbital fit results in a significantly different proper 
motion in right ascension. The orbital inclination is $51.3 ^{+4.6}_{-4.0}$\degr 
and results in a secondary mass of $0.37 \pm 0.03$\,M$_{\sun}$. The system's 
relative separation is $a_{rel}\simeq146$ mas and may be resolved by future high-resolution observations (e.g.\, 
\citealt{Mason:2001fk2}).

\subsection{HD\,158222}
The astrometric orbit is found with a significance of $99.3$\% ($2.7\sigma$), 
which is at the limit of a genuine detection. However, inspection of the 
confidence contours (Fig.~\ref{fig:contour}) shows that the inclination and 
$\Omega$ are well-constrained, and so the orbital solution is valid. The 
orbital inclination is $62.4 ^{+5.6}_{-5.1}$\degr and produces a secondary 
mass of $0.38 \pm 0.02$\,M$_{\sun}$.

\subsection{HD\,140667}
\begin{figure*}
\begin{center} 
\includegraphics[width= 0.31\linewidth, trim= 0 0.4cm 0 0.5cm]{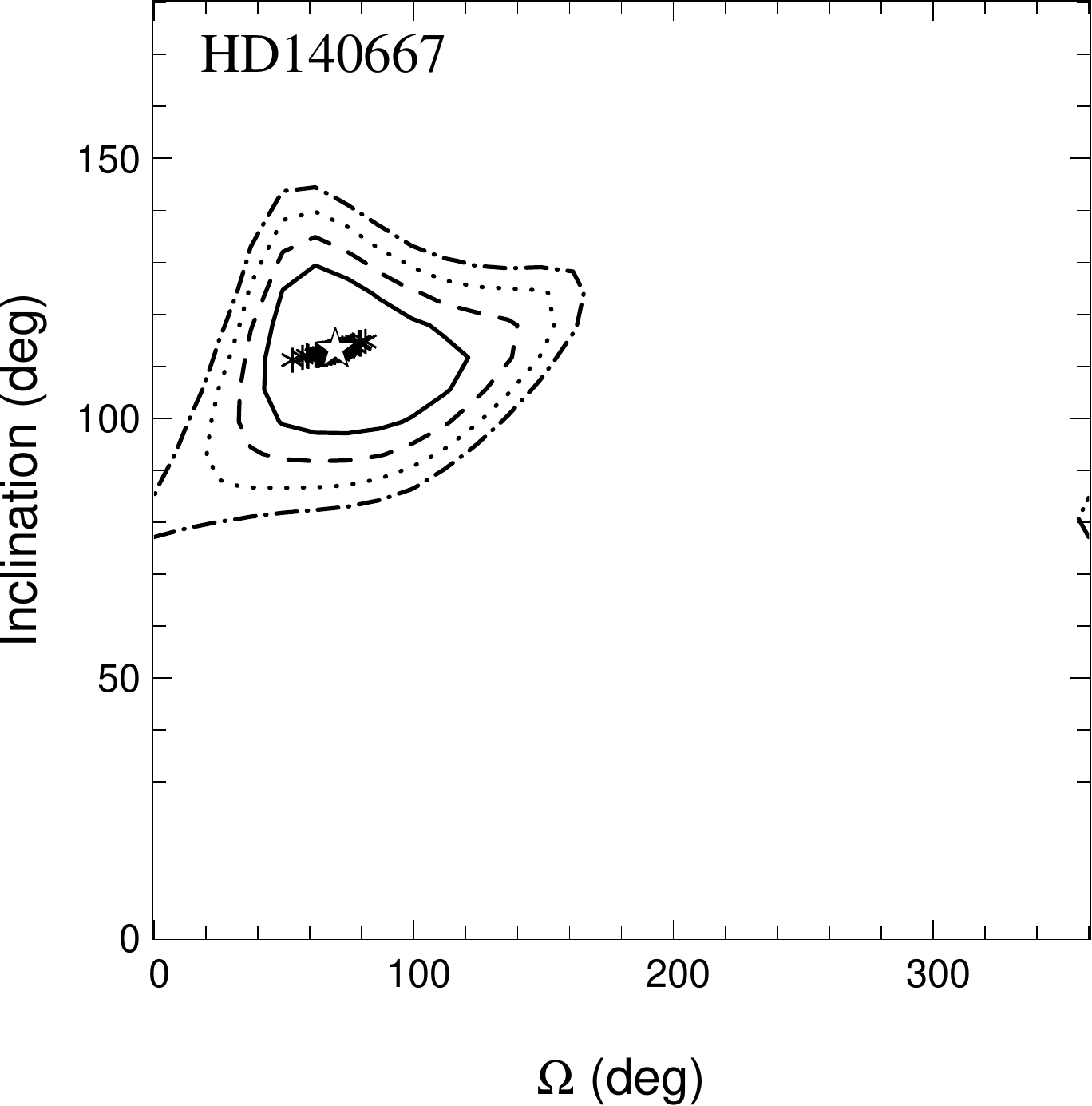}\hspace{3mm} 
\includegraphics[width= 0.31\linewidth, trim= 0 0.4cm 0 0.5cm]{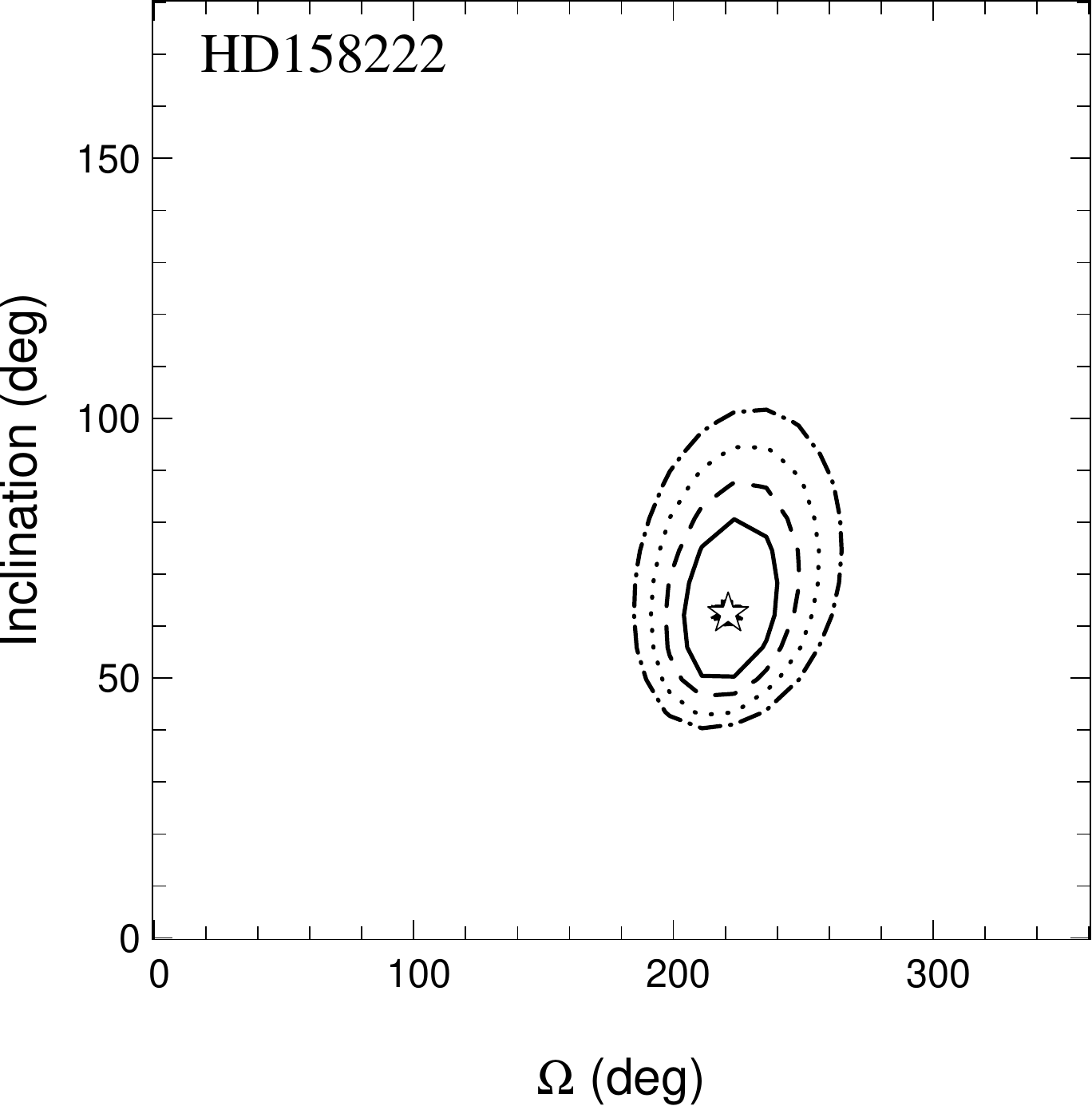}\hspace{3mm}  
\includegraphics[width= 0.31\linewidth, trim= 0 0.4cm 0 0.5cm]{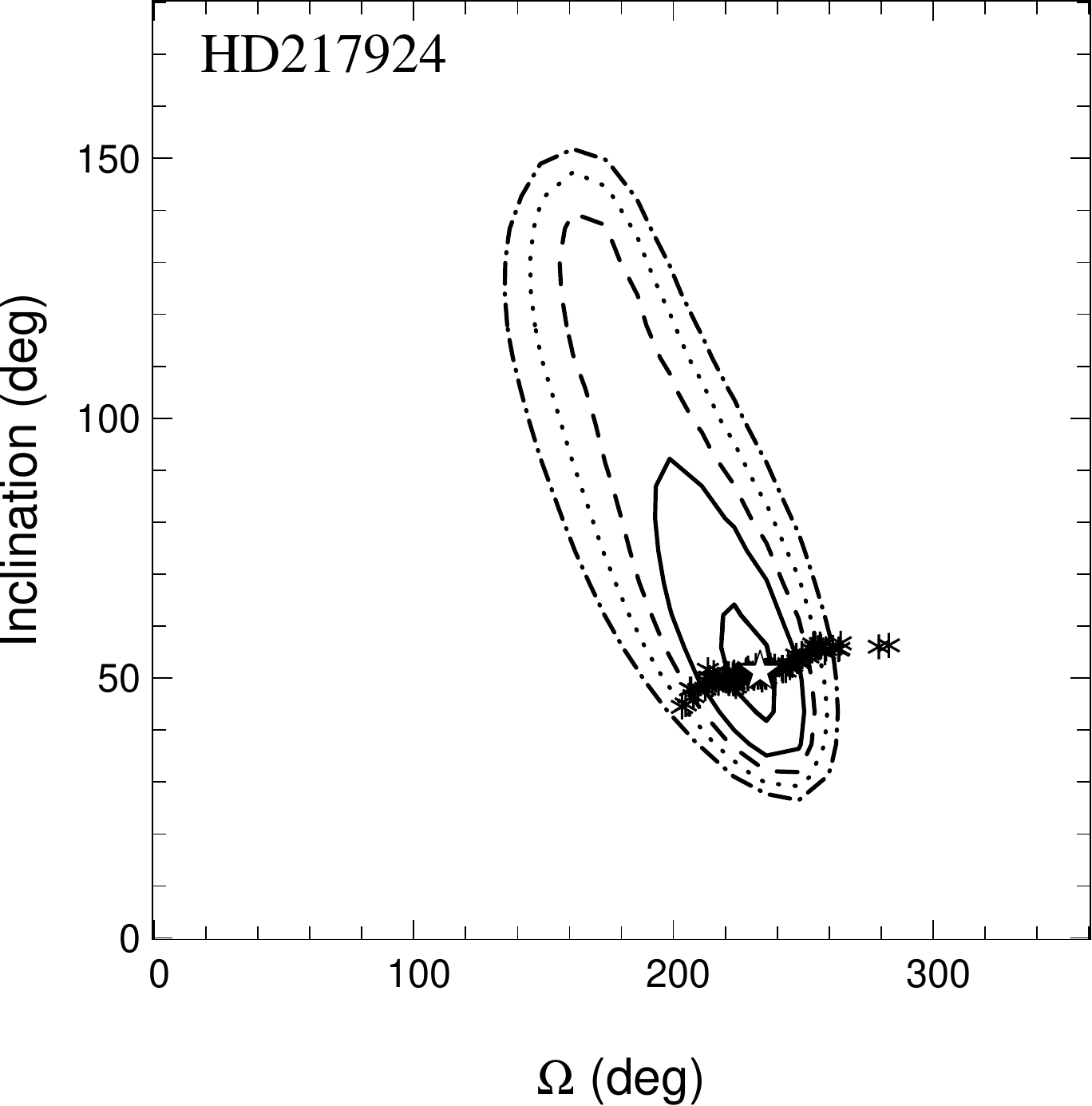} 
\caption{Joint confidence contours on the $i$-$\Omega$-grid for HD\,140667 (\emph{left}), HD\,158222 (\emph{centre}), HD\,217924 (\emph{right}). The contour lines 
correspond to confidences at 1-$\sigma$ (solid), 2-$\sigma$ (dashed), 
3-$\sigma$ (dotted), and 4-$\sigma$ (dash-dotted) level. The crosses indicate 
the position of the best non-linear adjustment solution for each of the 100 
Monte Carlo samples of spectroscopic parameters and the star corresponds to 
the adopted orbit.} 
\label{fig:contour}
\end{center} \end{figure*}

\begin{figure*}
\begin{center} 
\includegraphics[width= 0.31\linewidth, trim= 0 0.4cm 0 0.5cm]{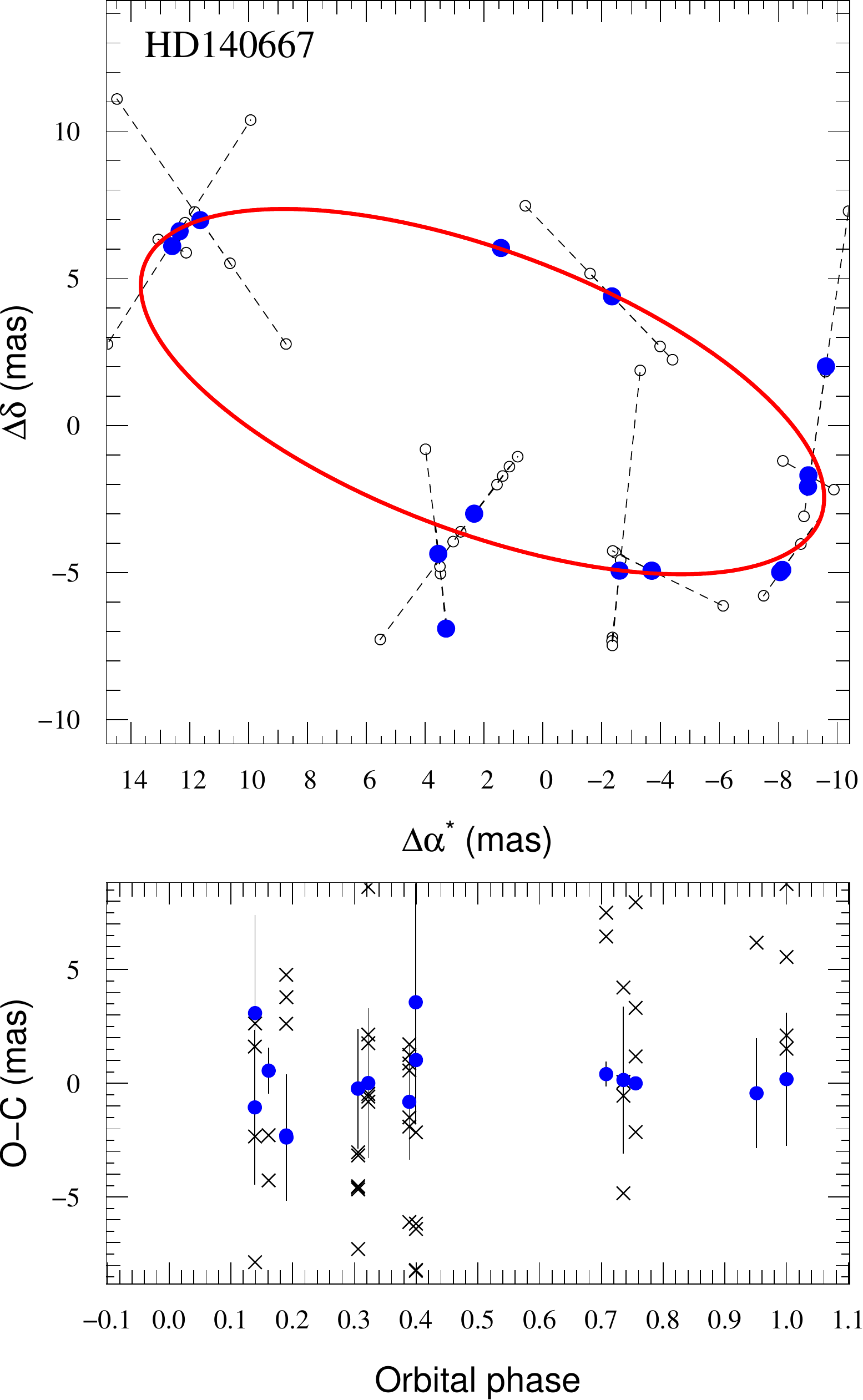}\hspace{3mm}
\includegraphics[width= 0.31\linewidth, trim= 0 0.4cm 0 0.5cm]{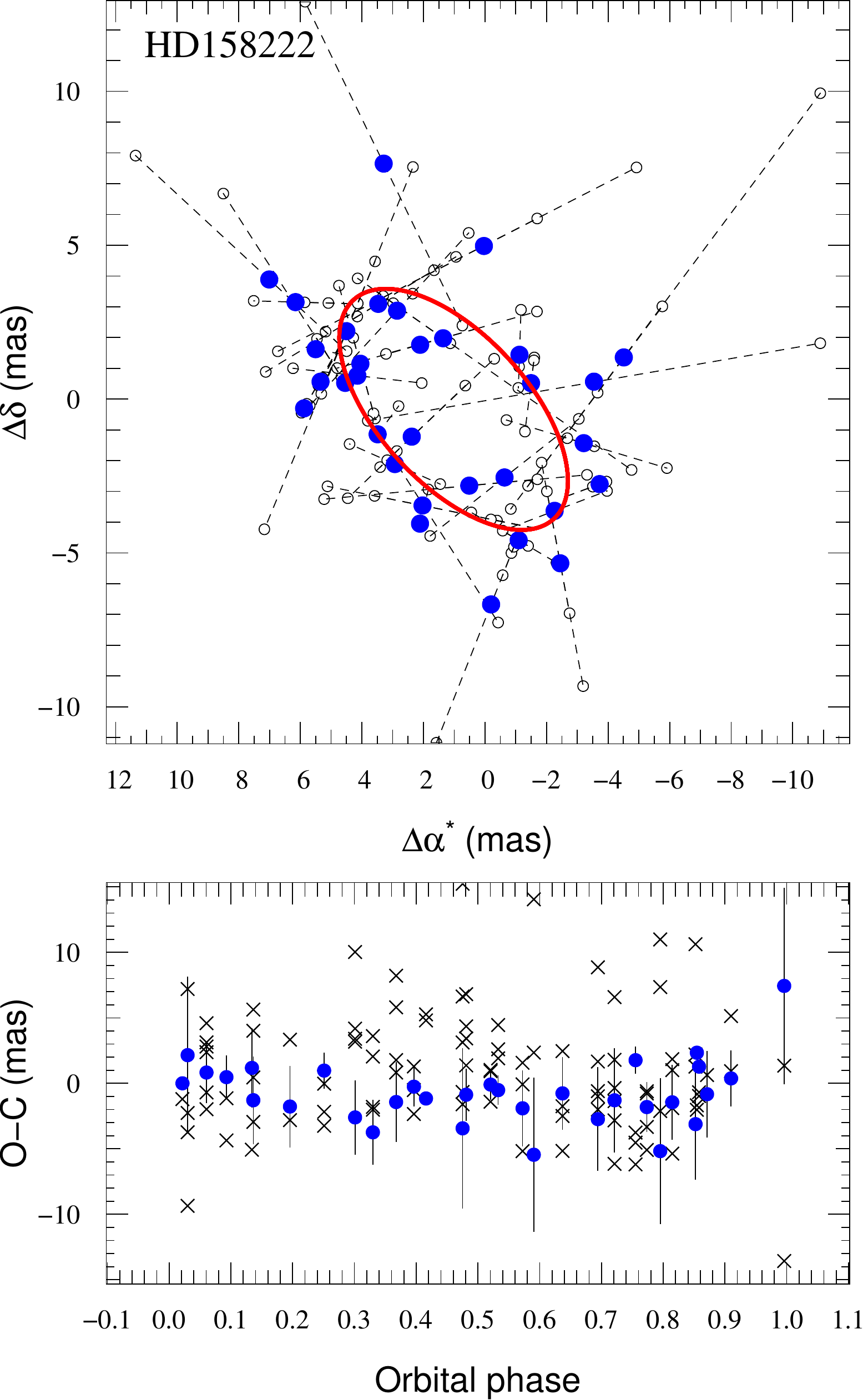}\hspace{3mm}
\includegraphics[width= 0.31\linewidth, trim= 0 0.4cm 0 0.5cm]{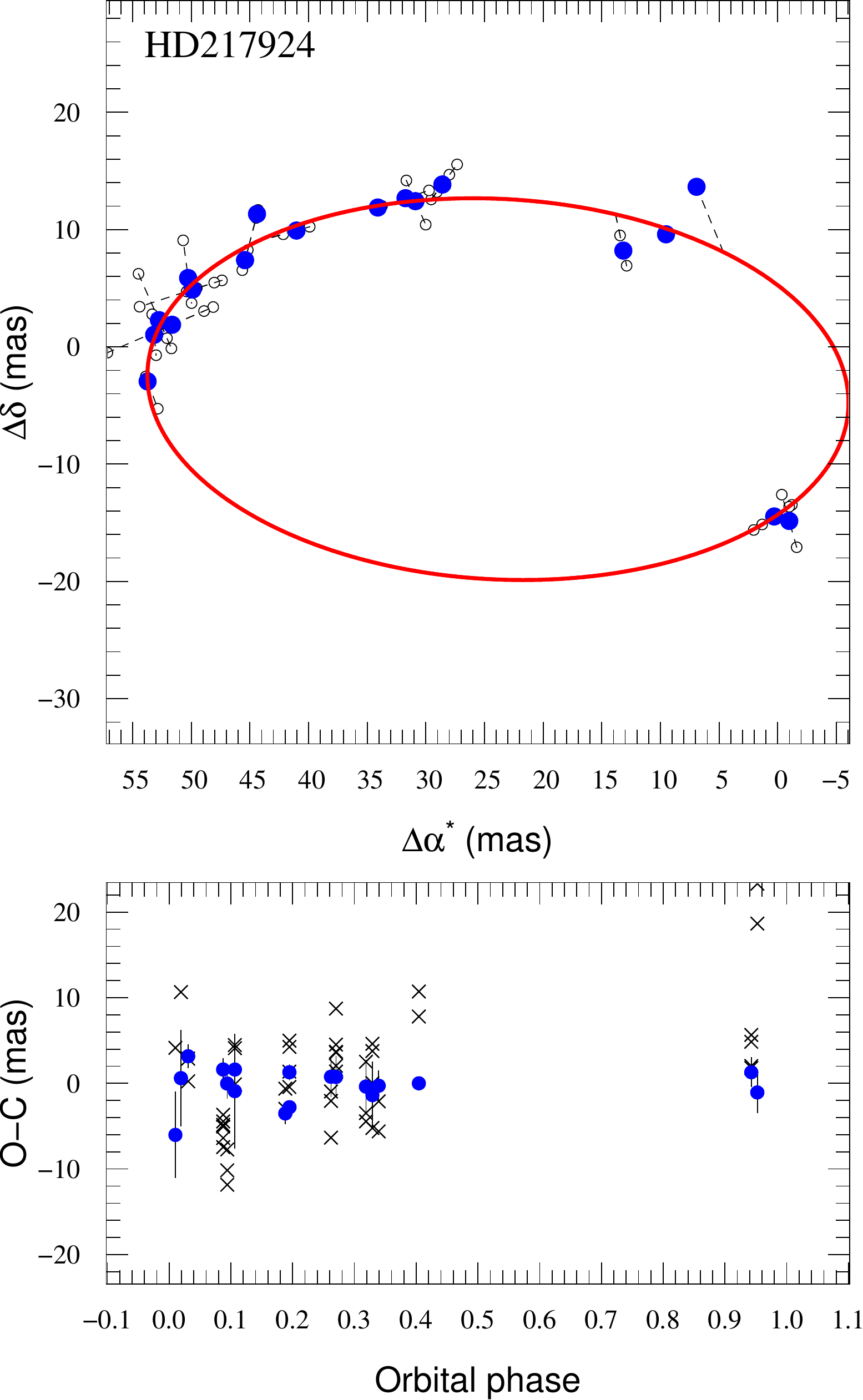}
\caption{Visualisation of the barycentric orbits. \emph{Top panels:} The sky-projected
orbits of HD\,140667 (\emph{left}), HD\,158222 (\emph{centre}), HD\,217924 (\emph{right}) are oriented clockwise, counter-clockwise, and counter-clockwise, respectively. North is up and east is left. 
The solid red line shows the orbital solution and open circles mark the 
individual Hipparcos measurements. \emph{Bottom panels:} The corresponding O-C residuals for the 
normal points of the orbital solution (filled blue circles) and of the 
standard 5-parameter model without companion (black crosses).} 
\label{fig:orbits}
\end{center} \end{figure*}

The permutation test yields a significance of only 12\% for the orbit of 
HD\,140667, which results in a non-detection of orbital motion on the basis of 
this criterion. However, the F-test probability is lower than for 
HD\,158222 indicating that orbital motion is present and is manifested 
in significantly reduced residuals. Inspection of the confidence contours 
(Fig.~\ref{fig:contour}) and of the orbit (Fig.~\ref{fig:orbits}) also 
strongly support a significant detection. We therefore adopt the formal 
solution of our analysis and claim that the orbit is constrained by 
Hipparcos astrometry. The orbital inclination is then $113.3 ^{+6.0}_{-6.5}$\degr and produces a secondary 
mass of $0.29 \pm 0.02$\,M$_{\sun}$. By including the orbital model, the parallax 
becomes significantly smaller by $\sim$2\,$\sigma$, thus the distance to 
the system is now larger. The revised parallax for HD\,140667 increases 
its luminosity $L/L_\sun$ from 0.874 to 1.107 and radius $R/R_\sun$ from 
0.91 to 1.03, thereby shifting its position in the H-R diagram 
(Fig.~\ref{fig:HR}) towards better agreement with evolutionary models.\\
We did not find an explanation for the failure of 
the permutation test, but we note that it is the only case that we have 
encountered so far in the analysis of more than 100 systems with identical 
methods, of which many are reported in \cite{Sahlmann:2011lr, Sahlmann:2011fk}, \cite{Diaz:2012fk}, and \cite{Marmier:2013fk}. For HD\,140667, we discarded the four IAD entries taken at satellite orbit number 1354, because those had uncertainties $>\!2\,\sigma_\Lambda$ and produced abnormally large residuals to the orbital fit (following \citealt{Sahlmann:2011fk}). The relatively small number of 48 effectively used IAD measurements is the same
as that used for HD\,217924 and cannot explain the failure of the permutation test.

\subsection{Companion mass limits for HD\,100167 and HD\,135991}
Even if evidence of orbital motion is not detected in the astrometric data, 
we can use the Hipparcos observations to set an upper limit to the companion 
mass by determining the minimum detectable astrometric signal $a_\mathrm{min}$ 
of the individual target. When the data cover at least one complete orbit, 
\cite{Sahlmann:2011fk, Sahlmann:2011lr} have shown that an astrometric 
signal-to-noise of $\mathrm{S/N} \gtrsim 6-7$ is required to obtain a 
detection at the $3\,\sigma$ level, where $\mathrm{S/N}=a\cdot 
(\sigma_{\Lambda}/\sqrt{N_\mathrm{Hip}})^{-1}$ and $a$ is the semi-major 
axis of the detected orbit. Using a conservative S/N-limit of 8, we derive 
the upper companion mass limit $M_{2,up-lim}$ as the companion mass which 
introduces the astrometric signal 
\begin{equation}\label{ }
a_\mathrm{min} = 8 \frac{ \sigma_{\Lambda} } { \sqrt{N_\mathrm{Hip}}} \left( 1-e^2 \right),
\end{equation}
where the factor $1-e^2$ accounts for the most unfavourable case of 
$i=90\degr$ and $\Omega=90\degr$ in which the astrometric signal is given 
by the semi-minor axis of the orbit. This criterion sets mass upper limits 
of 0.69\,$M_\sun$ and 0.59\,$M_\sun$ to the companions of HD\ 100167 and 
HD\ 135991, respectively.


\section{Discussion}
\begin{figure}
\begin{center} 
\includegraphics[width= \linewidth, trim= 0 1.5cm 0 1.5cm]{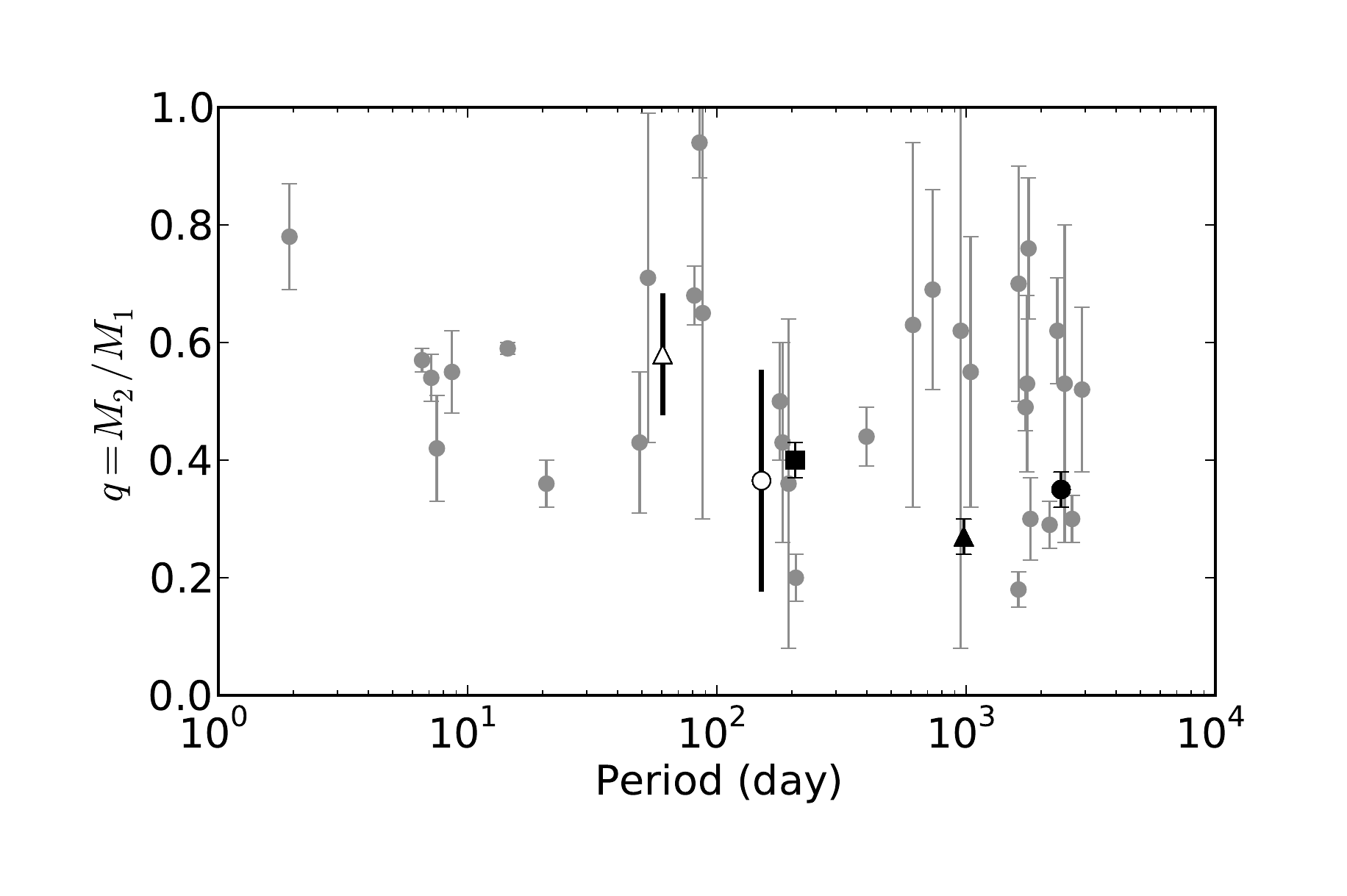}
\caption{The mass ratios as a function of period for the five SB1s are 
shown in black (symbols as in Fig.~\ref{fig:HR}) and the SB2 mass ratios 
from \cite{Mazeh:2003fe} are shown in grey.\vspace{-0.5cm}}
\label{fig:massratios}
\end{center} \end{figure}

The resulting mass ratios for the five SB1s are listed in Table~\ref{tab:44}. 
We adopted a systematic 10~\% model uncertainty for the primary mass estimate, 
which is propagated to the mass ratio uncertainty and is larger than the 
formal uncertainty obtained from the astrometric orbit fitting. For the 
two systems without detected orbits, we give the acceptable range in mass 
ratio. In Fig.~\ref{fig:massratios}, we compare our results with the sample 
of 32 SB2s with determined mass ratio from \cite{Mazeh:2003fe}. Although the SB1 primaries are on average $\sim$1.4 times more massive than the primaries of the SB2s examined by \cite{Mazeh:2003fe}, their mass ratios appear to be consistent with those of the lower mass primaries. If the SB1s studied here can be converted to SB2s using infrared spectroscopy, the detection of the astrometric orbit will yield model-independent dynamical masses for all components.

\begin{table}[h!] 
\caption{Mass ratio constraints. Uncertainties in brackets originate in the systematic primary mass uncertainty of 10~\%.}
\label{tab:44} 
\centering  
\begin{tabular}{c c c} 
\hline\hline %
Primary        &$M_1$           & $q$ \\  
            &($M_{\sun}$) &         \\  
\hline 
\object{HD\,100167} & $1.01 \left[\pm 0.10\right]$ & $0.48-0.68$ \vspace{1mm}\\ 
\object{HD\,135991} & $1.07 \left[\pm 0.11\right]$ & $0.18-0.55$ \vspace{1mm}\\ 
\object{HD\,140667} & $1.04 \left[\pm 0.10\right]$ & $0.27^{+0.01}_{-0.02} \left[^{+0.03}_{-0.02}\right]$\vspace{1mm} \\ 
\object{HD\,158222} & $0.94 \left[\pm 0.09\right]$ & $0.40^{+0.02}_{-0.02} \left[^{+0.04}_{-0.04}\right]$ \vspace{1mm}\\ 
\object{HD\,217924} & $1.05 \left[\pm 0.11\right]$ & $0.35^{+0.03}_{-0.03} \left[^{+0.04}_{-0.03}\right]$\vspace{1mm} \\ 
\hline
\end{tabular} 
\end{table}

\begin{acknowledgements}
J.S. thanks the Swiss National Science Foundation for supporting this research and kindly acknowledges support as a visitor at the Centro de Astrobiolog\'ia in Villanueva de la Ca\~nada (Madrid). The work of F.C.F. is partially supported by the state of Tennessee through its Centers of Excellence program.
\end{acknowledgements}
\bibliographystyle{aa} 
\bibliography{/Users/sahlmann/astro/papers} 
\end{document}